\documentclass{article}

\usepackage{PRIMEarxiv}

\usepackage[utf8]{inputenc} % allow utf-8 input
\usepackage[T1]{fontenc}    % use 8-bit T1 fonts
\usepackage{hyperref}       % hyperlinks
\usepackage{url}            % simple URL typesetting
\usepackage{booktabs}       % professional-quality tables
\usepackage{amsfonts}       % blackboard math symbols
\usepackage{nicefrac}       % compact symbols for 1/2, etc.
\usepackage{microtype}      % microtypography
\usepackage{lipsum}
\usepackage{fancyhdr}       % header
\usepackage{graphicx}       % graphics
\graphicspath{{media/}}     % organize your images and other figures under media/ folder

\DeclareUnicodeCharacter{202F}{}

\title{ Perceived Fairness And The Machine Learning Development Process: Concept Scale Development
}

\author{
  Anoop Mishra \\
  University of Nebraska \\
  6001 Dodge Street\\
  Omaha, Nebraska \\
  USA - 68182\\
  \texttt{amishra@unomaha.edu} \\
   \And
   Deepak Khazanchi \\
   University of Nebraska \\
   6001 Dodge Street\\
   Omaha, Nebraska \\
   USA - 68182\\
   \texttt{khazanchi@unomaha.edu} \\
}

\begin{document}
\maketitle

\begin{abstract}
In machine learning (ML) applications, unfairness is triggered due to bias in the data, the data curation process, erroneous assumptions, and implicit bias rendered during the development process. It is also well-accepted by researchers that fairness in ML application development is highly subjective, with a lack of clarity of what it means from an ML development and implementation perspective. Thus, in this research, we investigate and formalize the notion of the perceived fairness of ML development from a sociotechnical lens. Our goal in this research is to understand the characteristics of perceived fairness in ML applications. We address this research goal using a three-pronged strategy: 1) conducting virtual focus groups with ML developers, 2) reviewing existing literature on fairness in ML, and 3) incorporating aspects of justice theory relating to procedural and distributive justice. Based on our theoretical exposition, we propose operational attributes of perceived fairness to be transparency, accountability, and representativeness. These are described in terms of multiple concepts that comprise each dimension of perceived fairness. We use this operationalization to empirically validate the notion of perceived fairness of machine learning (ML) applications from both the ML practioners and users perspectives. The multidimensional framework for perceived fairness offers a comprehensive understanding of perceived fairness, which can guide the creation of fair ML systems with positive implications for society and businesses.
\end{abstract}

% keywords can be removed
\keywords{machine learning \and perceived fairness\and procedural justice theory}

\section{Proposed Definition And Attributes of Perceived Fairness}

\begin{figure}[t]
\centering
\includegraphics[width=0.8\columnwidth]{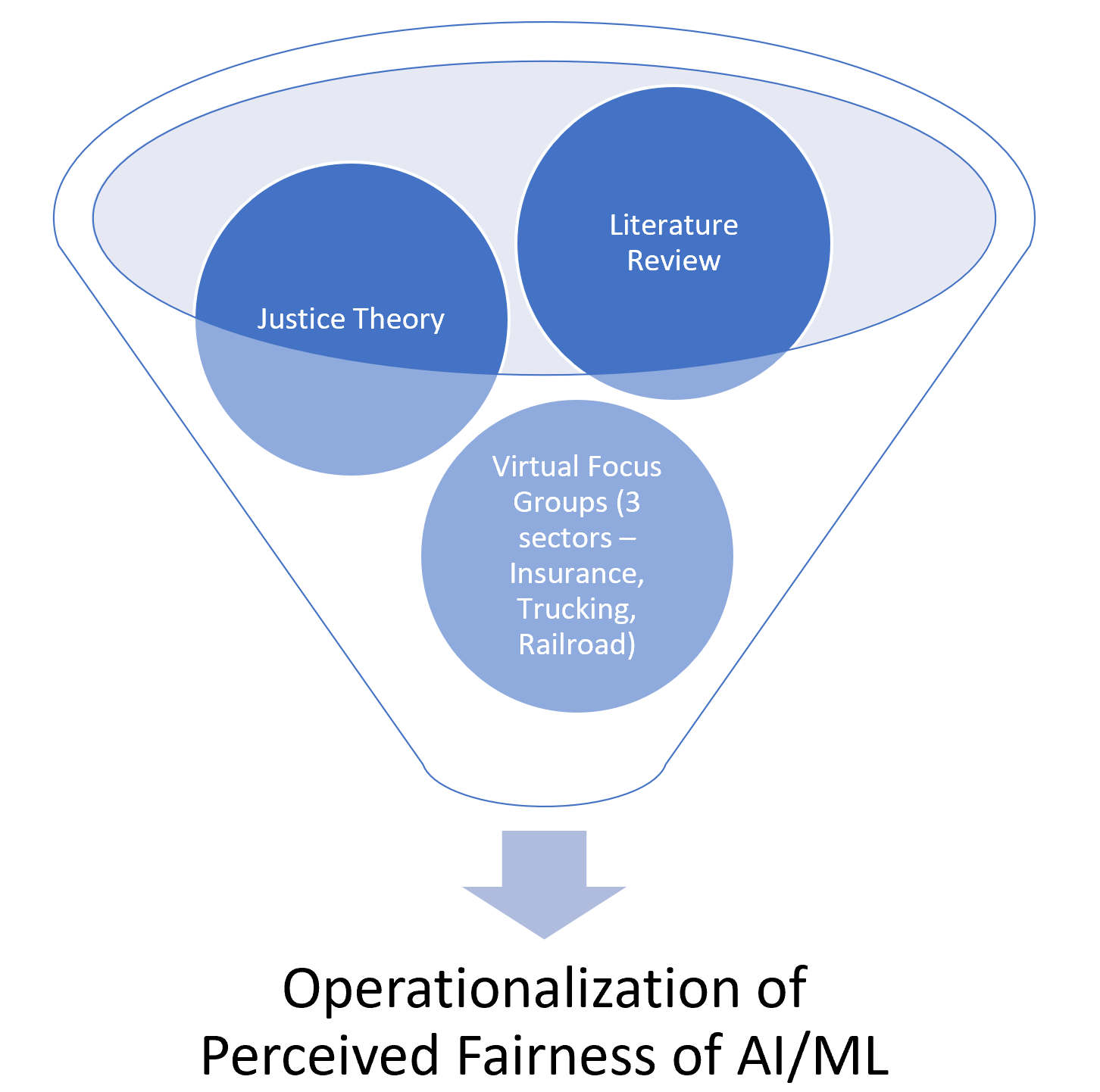}
\caption{Proposing perceived fairness as construct}
\label{funnel}
\end{figure}

This research study conducted virtual focus groups with developers, reviewed prior literature on fairness, and integrated notions of organizational justice theory to address the research question. Figure \ref{funnel} demonstrates the proposal for operationalizing the perceived fairness as a construct. This section discusses the findings of virtual focus group, operational definition, proposed attributes, and sub-attributes of perceived fairness.

%\subsection{Face Validity: Virtual Focus Groups}
\subsection{Virtual Focus Groups}
A focus group in research is a group discussion of people with similar characteristics who share experiences and discuss to generate data \cite{khazanchi2006patterns,kitzinger1995qualitative}. Focus group discussions are utilized as a qualitative approach to assessing an in-depth understanding of social issues \cite{o2018use, khazanchi2006patterns}. This research study uses focus groups to explore ML developer’s perceptions of fairness. The participants targeted for the focus groups are ML developers, data scientists, and engineers from the industry who participate in designing and developing ML applications. A total of 20 participants from three different industry organizations signed up for the focus group, but only 9 participants showed up. The research has Institutional Review Board (IRB) approval for conducting the focus groups. The virtual focus group was designed to develop an understanding of perceived fairness and its attributes among developers. This research study utilized MIRO as a brainstorming tool and Zoom to conduct the focus groups. MIRO is an online visual platform where teams can connect, collaborate, create, and brainstorm together (see \url{https://miro.com/}). All sessions are conducted synchronously. Each company, along with its participants, has one focus group. Each focus group is given a 75-minute window to participate. Participants from each company were provided with their own unique session ID on MIRO. More details about virtual focus group design and questions can be found in the Appendix section. In this section, we present the findings from the focus group data collected from ML developers. \par

The focus groups suggest that participant's ideas and discussions are influenced by their personal experience, knowledge base, and practice gained through developing ML applications. An inductive approach using thematic analysis and topic modeling using Latent Dirichlet Allocation (LDA) assisted in deriving themes from the developer's discussion on focus groups. These themes are bias mitigation, data, model design, model validity, business rules, and user interaction, which describe the developer's perceived fairness. \textbf{Developers discussion on \textit{"fairness"}}: Based on the findings of our study and the discussion above, we conclude that the developer’s perceived fairness comprises the complete ML process, including privacy, ethics, the intention of ML development, business constraints and goals, explainability to users, and user’s usability. Interestingly, one of the developers claims that fairness in machine learning is a subjective term and that the evaluation of ML models must include the ML pipeline process.

%The supportive evidence from focus groups in table \ref{Procedural} shows the transcripts from focus group discussions.

\begin{table*}[htp]
\centering
\caption{Relationship between themes of fairness and procedural (justice) fairness components from literature}
\label{Procedural}
%\begin{minipage}{\columnwidth}
\begin{center}
\begin{tabular}{|p{1in}|p{1.3in}|p{1.5in}|p{1.5in}|}
\hline
\textbf{Themes} & \textbf{Lee et al. 2019 } & \textbf{Rueda 2022 } & \textbf{Leventhal 1980; Morse et al 2020 }\\
\hline
Bias Mitigation
&
Control  
&
Avoidance of bias, Accountability 
&
Bias suppression 
\\
\hline
Data
&
Control, Transparency
&
Avoidance of bias 
&
Consistency, Representiveness 
\\
\hline
Model Design  
&
Transparency 
&
Transparency 
&
Correctability 
\\
\hline
Model Validity 
&
Transparency, Control, Principle 
&
Transparency, Accountability 
&
Accuracy, Correctability 
\\
\hline
Business Rules
&
Transparency, 
&
Avoidance of bias, Accountability 
&
Ethicality 
\\
\hline
Users Interaction
&
Transparency, Control, Principle 
&
Accountability 
&
Ethicality 
\\
\hline
\end{tabular}
\end{center}
\end{table*}

\subsection{Proposed attributes of perceived fairness} 

In the related works section, the components of procedural fairness from organization justice theory discovered in literature are reviewed. Lee \textit{et al.} 2019 describes procedural fairness using transparency, control, and principle \cite{lee2019procedural}. As per Lee \textit{et al.} 2019, transparency is the rules of the decision-maker that are perceived as fair and warranted, including an explanation of decision outcomes and information representativeness. Control is described as the degree of control over the decision that individuals receive, and principle is defined as demonstrations of consistency, competency, benevolence, and voice. Rueda 2022 explains procedural fairness as avoidance of bias, accountability, and transparency in medical scenarios. Rueda 2022 defines transparency as the procedure that explains ML algorithms working and processing that lead to the outcome. Accountability is also related to the robustness of the model, and avoidance of bias describes not including attributes that can cause unfavorable decisions \cite{rueda2022just}. Morse \textit{et al.} 2021 discuss the components of procedural fairness proposed by Leventhal 1980 as bias impression, consistency, representativeness, correctability, accuracy, and ethicality. Consistency defines the uniformity of decision procedures across people and time, accuracy is the measure of validity and high-quality information, ethicality describes practicing moral standards and values, representativeness describes proper population representation, bias suppression subjects to prevent favoritism by the decision maker, and lastly, correctability are approaches to correct flawed decisions \cite{Leventhal1980,morse2021ends}. Table \ref{Procedural} shows an association of themes describing the developer's perceived fairness (from virtual focus groups) with the components of procedural fairness proposed by Lee \textit{et al.} 2019, Rueda 2022, and Leventhal 1980. These associations are proposed by the union of the procedural fairness component's description from the literature discussed and the ML developer's discussion in the focus groups. For example, the theme "Data" aligns with Lee's \textit{et al.} 2019 transparency and control because of information representativeness and its impact on decisions in the data-driven process. It also aligns with Rueda's 2022 avoidance of bias for fair decision-making and Leventhal's 1980 consistency and representativeness for uniform decision-making across people and time. Thus, we conclude that the association in table \ref{Procedural} illustrates that aspects of organizational justice theory, i.e., procedural fairness and distributed fairness, can explain the developer’s and user's perception of ML fairness. \par

\begin{figure}[t]
\centering
\includegraphics[width=0.8\columnwidth]{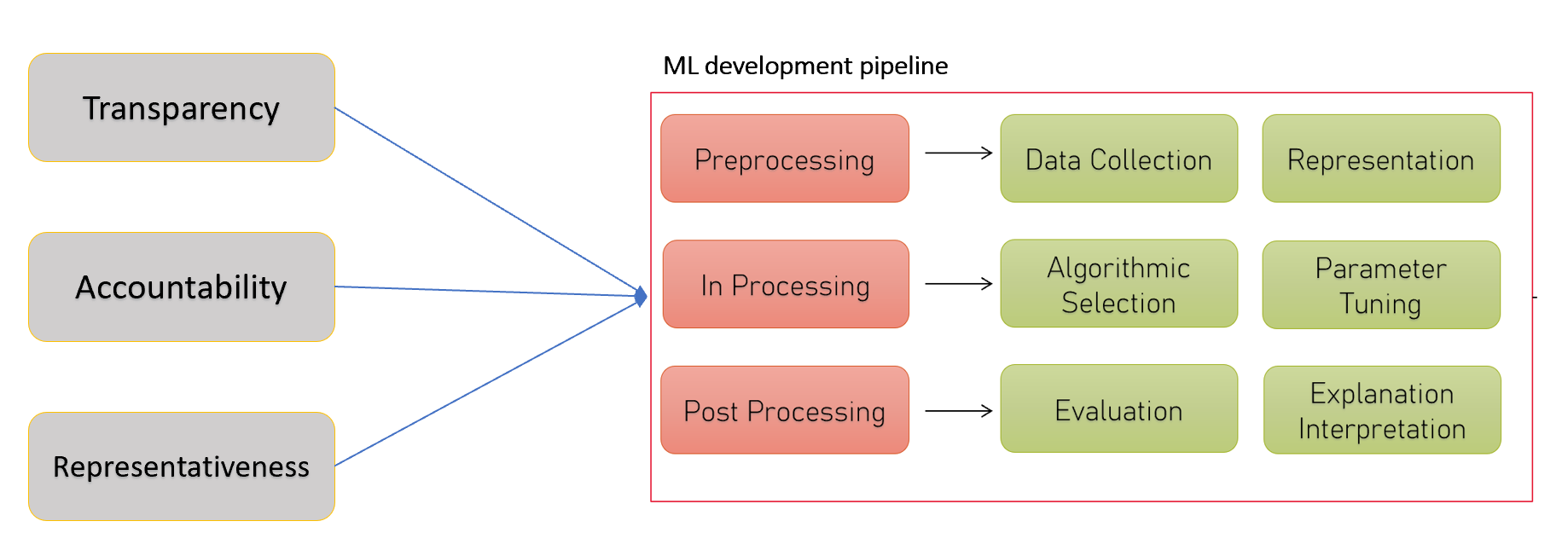}
%includegraphics[width=0.8\textwidth]{AnonymousSubmission/LaTeX/funnel.png} % Reduce the figure size so that it is slightly narrower than the column.
\caption{A conceptual framework with operational attributed of perceived fairness in ML applications}
\label{attributes}
\end{figure}

The construct \textit{perceived fairness} incorporates attributes to evaluate the beliefs of ML practitioners and users regarding fairness. Figure \ref{funnel} describes the conceptual development of perceived fairness utilizing notions of justice theory, systematic literature review, and virtual focus groups. The three attributes proposed for perceived fairness are transparency, accountability, and representativeness. These attributes consist of eleven sub-attributes. These three attributes, along with their sub-attributes, will be operationalized for assessing the perceived fairness of the ML application both from the developer's and the user's perspective. The conceptual framework, including these operational attributes, is shown in figure \ref{attributes}. The definitions of attributes and sub-attributes are formulated to encompass both the principles of procedural and distributed fairness. The comprehensive description of the proposed operationalized definition of three attributes, along with their respective measures, is described below. \par

\textbf{Transparency} is the degree to which the ML process, data, and functions are traceable, explainable, and have fidelity in an ML application. 
\begin{enumerate}
    \item \textbf{Traceability} is the extent to which the ML process at each step is trackable.
    \item \textbf{Explainability} or “interpretability” is the extent to which the ML model and its output are explained in a way that “makes sense” to a human being at an acceptable level. 
    \item \textbf{Fidelity} is the ability to ensure and demonstrate the validity of the ML process based on explanations. (describes how well the explanation of a machine learning model approximates the prediction of the black box model)
\end{enumerate}

\textbf{Accountability} is the degree to which the ML process incorporates accuracy, correctability, consistency, ethicality, and governance.
\begin{enumerate}
    \item \textbf{Accuracy} is the ability to ensure and demonstrate the validity and precision of the ML process.
    \item \textbf{Consistency} is the ability to ensure and demonstrate the stability and reliability of the ML process.
    \item \textbf{Correctability} is the ability to detect and correct any errors, flaws, or biases in the ML process.
    \item \textbf{Ethicality} is the ability to ensure and demonstrate the morality and responsibility of the ML process.
    \item \textbf{Governance} is the ability to systematically track, measure, manage, and monitor the ML development and deployment process while holding its developers and users accountable. 
\end{enumerate}

\textbf{Representativeness} is defined as the extent to which the data used in the ML process for training and testing is a true representation of the population to which the model is being applied. 
\begin{enumerate}
    \item \textbf{ML Development process} is the extent to which the ML process considers and incorporates the diversity and variability of the population and the context to which the model is being applied. 
    \item \textbf{Training of ML models} is the extent to which the ML models learn and capture the patterns and relationships of the target population and the context accurately and fairly.
    \item \textbf{Interpretation of ML outputs} is the extent to which the ML outputs are meaningful and relevant for the target population and the context.
    
\end{enumerate}
%%%%%%%%%%%%%%

\subsection{Definition to construct 'perceived fairness'}
Based on the proposed attributes and from figure \ref{funnel} and \ref{attributes}, the construct is defined as- \textit{Perceived fairness of ML applications is described in terms of systems that are designed and built to be fair in processes, transparent in actions (explainable), have the opportunity for multiple voices to be integrated into their development, and are impartial to all users in their outcomes.}

%%%%%%%%%%%%%%%%%%%%%%%%%%%%%%%%%%%%%%%%%%%%%%%%%%%%%%%%%%%%%%%%%%%%%%%%%%%%%%%

\subsection{Future work}
This research study identified the research gap and proposed a research question to understand the human perception of fairness from a socio-technical lens utilizing organizational justice theory. Further, this study identifies the attributes and proposes their operational definitions and a conceptual framework that will help to assess the perception of humans (both developers and users) in ML applications. Further, the following can be potential future work utilizing the proposed framework of perceived fairness-
\begin{itemize}
    \item an instrument (like survey method) and test cases can be developed that can validate the proposed attributes
    \item a tool development (plugin) that can measure and assess the perceived fairness both from the ML developer’s and user’s perspective, utilizing proposed attributes for the ML process.
    \item a comparative study that can holistically measure the difference between developer's and user's perceptions of fairness in ML application
\end{itemize}

%%%%%%%%%%%%%%%%%%%%%%%%%%%%%%%%%%%%%%%%%%%%%%%%%%%%%%%%%%%%%%%%%%%%%%%%%%%%%%%

\section{Conclusions}

Unfairness in ML can arise due to biased data, flawed assumptions, implicit biases, and curation processes. Creating fair ML systems is crucial for responsible AI as ML grows more widespread. However, fairness in ML is highly subjective, with no consistent way to describe the fairness of ML system. This research study aimed to understand human-centric fairness from a socio-technical lens in all the phases of the ML development lifecycle. Thus, the study investigated the perception of fairness in the machine learning (ML) process using a three-way approach, including virtual focus groups with developers, reviewing prior work, and integrating justice theory. The outcome of this approach is a comprehensive understanding of perceived fairness in the ML process and ensuring its operationalization in the ML development cycle. We propose a conceptual framework defining perceived fairness as a multidimensional construct that can be operationalized by utilizing the proposed attributes, such as transparency, accountability, and representativeness, along with their sub-attributes operational definitions, in the ML development lifecycle. Utilizing the framework along with the operational attributes can help to assess and understand the perceived fairness to develop fair and ethical ML applications.

\section{Acknowledgments}
It acknowledges a support grant from the U.S. Army Corps of Engineers, Engineering Research and Development Center (ERDC) for this research (Grant Number: W912HZ21C0060 and W912HZ23C0005). It also acknowledges earlier publications in the series \cite{mishra2023assessing, mishra2023neurips, mishragraca2022, mishrabigxii2022, mishraproposal}.

\bibliographystyle{unsrtnat}
\bibliography{references}

\begin{thebibliography}{12}
\providecommand{\natexlab}[1]{#1}
\providecommand{\url}[1]{\texttt{#1}}
\expandafter\ifx\csname urlstyle\endcsname\relax
  \providecommand{\doi}[1]{doi: #1}\else
  \providecommand{\doi}{doi: \begingroup \urlstyle{rm}\Url}\fi

\bibitem[Khazanchi and Zigurs(2006)]{khazanchi2006patterns}
Deepak Khazanchi and Ilze Zigurs.
\newblock Patterns for effective management of virtual projects: Theory and evidence.
\newblock \emph{International Journal of e-Collaboration (IJeC)}, 2\penalty0 (3):\penalty0 25--49, 2006.

\bibitem[Kitzinger(1995)]{kitzinger1995qualitative}
Jenny Kitzinger.
\newblock Qualitative research: introducing focus groups.
\newblock \emph{Bmj}, 311\penalty0 (7000):\penalty0 299--302, 1995.

\bibitem[O.~Nyumba et~al.(2018)O.~Nyumba, Wilson, Derrick, and Mukherjee]{o2018use}
Tobias O.~Nyumba, Kerrie Wilson, Christina~J Derrick, and Nibedita Mukherjee.
\newblock The use of focus group discussion methodology: Insights from two decades of application in conservation.
\newblock \emph{Methods in Ecology and evolution}, 9\penalty0 (1):\penalty0 20--32, 2018.

\bibitem[Lee et~al.(2019)Lee, Jain, Cha, Ojha, and Kusbit]{lee2019procedural}
Min~Kyung Lee, Anuraag Jain, Hea~Jin Cha, Shashank Ojha, and Daniel Kusbit.
\newblock Procedural justice in algorithmic fairness: Leveraging transparency and outcome control for fair algorithmic mediation.
\newblock \emph{Proceedings of the ACM on Human-Computer Interaction}, 3\penalty0 (CSCW):\penalty0 1--26, 2019.

\bibitem[Rueda et~al.(2022)Rueda, Rodr{\'\i}guez, Jounou, Hortal-Carmona, Aus{\'\i}n, and Rodr{\'\i}guez-Arias]{rueda2022just}
Jon Rueda, Janet~Delgado Rodr{\'\i}guez, Iris~Parra Jounou, Joaqu{\'\i}n Hortal-Carmona, Txetxu Aus{\'\i}n, and David Rodr{\'\i}guez-Arias.
\newblock “just” accuracy? procedural fairness demands explainability in ai-based medical resource allocations.
\newblock \emph{AI \& society}, pages 1--12, 2022.

\bibitem[Leventhal(1980)]{Leventhal1980}
Gerald~S. Leventhal.
\newblock \emph{What Should Be Done with Equity Theory?}, pages 27--55.
\newblock Springer US, Boston, MA, 1980.
\newblock ISBN 978-1-4613-3087-5.
\newblock \doi{10.1007/978-1-4613-3087-5_2}.
\newblock URL \url{https://doi.org/10.1007/978-1-4613-3087-5_2}.

\bibitem[Morse et~al.(2021)Morse, Teodorescu, Awwad, and Kane]{morse2021ends}
Lily Morse, Mike Horia~M Teodorescu, Yazeed Awwad, and Gerald~C Kane.
\newblock Do the ends justify the means? variation in the distributive and procedural fairness of machine learning algorithms.
\newblock \emph{Journal of Business Ethics}, pages 1--13, 2021.

\bibitem[Mishra and Khazanchi(2023{\natexlab{a}})]{mishra2023assessing}
Anoop Mishra and Deepak Khazanchi.
\newblock Assessing perceived fairness from machine learning developer's perspective.
\newblock \emph{arXiv preprint arXiv:2304.03745}, 2023{\natexlab{a}}.

\bibitem[Mishra and Khazanchi(2023{\natexlab{b}})]{mishra2023neurips}
Anoop Mishra and Deepak Khazanchi.
\newblock Assessing perceived fairness in machine learning (ml) process: A conceptual framework.
\newblock In \emph{Algorithmic Fairness through the Lens of Time Workshop, NeurIPS}, 2023{\natexlab{b}}.

\bibitem[Mishra(2022{\natexlab{a}})]{mishragraca2022}
Anoop Mishra.
\newblock Perceived fairness from developer’s perspective in artificial intelligent systems.
\newblock 2022{\natexlab{a}}.

\bibitem[Mishra(2022{\natexlab{b}})]{mishrabigxii2022}
Anoop Mishra.
\newblock Perceived fairness from user's and developer's perspective in machine learning systems.
\newblock 2022{\natexlab{b}}.

\bibitem[Mishra(2024)]{mishraproposal}
Anoop Mishra.
\newblock \emph{{Perceived Fairness from User’s And Developer’s Perspective in Artificial Intelligent Systems, Ph.D. Proposal}}.
\newblock {Ph.D.} proposal, CIST, University of Nebraska at Omaha, Omaha, NE, 2024.

\end{thebibliography}

\end{document}